\documentclass[10pt,letterpaper]{article}
\usepackage[top=0.85in,left=2.75in,footskip=0.75in]{geometry}

\usepackage{amsmath,amssymb}

\usepackage{changepage}

\usepackage{textcomp,marvosym}

\usepackage{cite}

\usepackage{nameref,hyperref}

\usepackage[right]{lineno}

\usepackage[nopatch=eqnum]{microtype}
\DisableLigatures[f]{encoding = *, family = * }

\usepackage[table]{xcolor}

\usepackage{array}

\usepackage{algorithm}%
\usepackage{algorithmicx}%
\usepackage{algpseudocode}%

\usepackage{xcolor}

\newcolumntype{+}{!{\vrule width 2pt}}

\newlength\savedwidth

\newcommand\thickhline{\noalign{\global\savedwidth\arrayrulewidth\global\arrayrulewidth 2pt}%
\hline
\noalign{\global\arrayrulewidth\savedwidth}}

\raggedright
\usepackage[aboveskip=1pt,labelfont=bf,labelsep=period,justification=raggedright,singlelinecheck=off]{caption}

\makeatletter
\renewcommand{\@biblabel}[1]{\quad#1.}
\makeatother

\usepackage{lastpage,fancyhdr,graphicx}
\usepackage{epstopdf}
\fancyheadoffset[L]{2.25in}
\fancyfootoffset[L]{2.25in}
\begin{document}
\vspace*{0.2in}

\begin{flushleft}
{\Large
\textbf\newline{A step toward a reinforcement learning \textit{de novo} genome assembler} 
}
\newline
\\
Kleber Padovani\textsuperscript{1*\Yinyang\ddag\textcurrency\dag\textpilcrow\textsection},
Roberto Xavier\textsuperscript{2\Yinyang\textcurrency\textpilcrow\textsection},
Rafael Cabral Borges\textsuperscript{2\textpilcrow\textsection},
André Carlos Carvalho\textsuperscript{3\textcurrency\dag\textpilcrow},
Anna Reali\textsuperscript{4\textcurrency\dag\textpilcrow},
Annie Chateau\textsuperscript{5\dag\textpilcrow},
Ronnie Alves\textsuperscript{2,6\Yinyang\ddag\textcurrency\dag\textpilcrow}
\\
\bigskip
\textbf{1} Center for Higher Studies of Itacoatiara, University of the State of Amazonas, Itacoatiara, Amazonas, Brazil
\\
\textbf{2} Sustainable Development, Vale Technology Institute, Belém, Pará, Brazil
\\
\textbf{3} Institute of Mathematics and Computer Sciences, University of São Paulo, São Carlos, São Paulo, Brazil
\\
\textbf{4} Polytechnic School, University of São Paulo, São Paulo, São Paulo, Brazil
\\
\textbf{5} LIRMM, University of Montpellier, CNRS, Montpellier, France
\\
\textbf{6} Institute of Natural Sciences, Federal University of Pará, Belém, Pará Brazil
\\
\bigskip

%
%
\Yinyang Role(s): Conceptualization, Data Curation, Investigation

\ddag Role(s): Formal Analysis, Project Administration.

\textcurrency Role(s): Methodology 

\dag Role(s): Validation

\textsection Writing – Original Draft Preparation

\textpilcrow Role(s): Writing – Review \& Editing

* kleber.padovani@gmail.com (KP)

\end{flushleft}
\section*{Abstract}
\emph{De novo} genome assembly is a relevant but computationally complex task in genomics. Although \emph{de novo} assemblers have been used successfully in several genomics projects, there is still no `best assembler', and the choice and setup of assemblers still rely on bioinformatics experts. Thus, as with other computationally complex problems,  machine learning may emerge as an alternative (or complementary) way for developing more accurate and automated assemblers. Reinforcement learning has proven promising for solving complex activities without supervision -- such games -- and there is a pressing need to understand the limits of this approach to `real' problems, such as the DFA problem. This study aimed to shed light on the application of machine learning, using reinforcement learning (RL), in genome assembly. We expanded upon the sole previous approach found in the literature to solve this problem by carefully exploring the learning aspects of the proposed intelligent agent, which uses the Q-learning algorithm, and we provided insights for the next steps of automated genome assembly development. We improved the reward system and optimized the exploration of the state space based on pruning and in collaboration with evolutionary computing. We tested the new approaches on 23 new larger environments, which are all available on the internet. Our results suggest consistent performance progress; however, we also found limitations, especially concerning the high dimensionality of state and action spaces. Finally, we discuss paths for achieving efficient and automated genome assembly in real scenarios considering successful RL applications --- including deep reinforcement learning.



\section*{Introduction}

The genome of an organism is the sequence of all nucleotides from its DNA molecules. Each isolated nucleotide represents no relevant biological information, and within the organism's genome there are species genes, that define species traits and behaviors (e.g., eye color)~\cite{Portin2017}. A single DNA fragment cannot represent the complete information from a gene, and genome assembly is the computational task used to order the sequenced DNA fragments (i.e., \emph{reads}) and reconstruct the original DNA sequence~\cite{Heather2016}. The size and number of \emph{reads} directly impact the complexity of the assembly process, and illuminating this bottleneck problem has become an important bioinformatics problem for producing a fast, automated and reliable solution.

Genome assemblers adopt comparative and/or \emph{de novo} strategies. A comparative strategy (unlike \emph{de novo} strategies) requires a reference genome~\cite{Ji2017}. \emph{De novo} strategies are particularly important given that only a small number of reference genomes are currently available~\cite{Wong2020}. However, this approach is a highly complex combinatorial problem that falls into the theoretically intractable class of computational problems (NP-hard)~\cite{Medvedev2007}. \emph{De novo} assemblers (commonly based on heuristics and graphs) can produce acceptable solutions, but require specific bioinformatics knowledge for configuration and parameter setting, and optimal results are not guaranteed~\cite{Gurevich2013}.


Genome assembly is currently not a fully solved problem. Nevertheless, few approaches have been applied machine learning trying to achieve better solutions for the assembly problem~\cite{PadovanideSouza2018, Yassine2023}). With current access to increased processing and storage power, machine learning application started to increase and good results have been reported~\cite{LeCun2019}. This increase also allowed the return of reinforcement learning application for these problems~\cite{Botvinick2019}.

Reinforcement learning (RL) is a basic machine learning paradigm in which intelligent agents take actions in a task environment. Ideally, this agent solves the task when it is able to learn how to maximize the rewards from its actions~\cite{sutton2018}. Such low use of RL in real-world problems has also been observed in the particular instance of genome assembly, where few studies was found, ~\cite{bocicor2011, PadovanideSouza2018, xavier2020, Karami2023}. 

The seminal approach~\cite{bocicor2011} proposes an episodic trained agent (i.e., whose training has been divided into episodes) applying the Q-learning reinforcement learning algorithm, which allows the agent to learn through the consequences (positive or negative rewards) it receives after taking actions. Obtaining intelligent and trained agents by RL using the seminal approach is important because it could eliminate the need for human specialists.

The Q-learning algorithm requires a Markov decision process definition with established parameters of states and actions, together with a reward system to be achieved by the agent at each action in every state~\cite{sutton2018}. The problem was then modeled through a state space capable of representing all possible \emph{read} permutations, with one action for each \emph{read} in each state~\cite{bocicor2011}. Following these definitions, from graph theory, the proposed state space for $n$ \emph{reads} can be visualized as a complete $n$-ary tree, with height equal to $n$, as the set of states presents one initial state and forms a connected and acyclic graph~\cite{cormen2009}. The number of existing states in the state space is represented by Eq \ref{eq:04}.

\begin{equation}
\label{eq:04}
\textrm{number of states}=\frac{n^{n+1}-1}{n-1}
\end{equation}

\begin{figure}[!h]
\centering
\includegraphics[width=6cm,keepaspectratio]{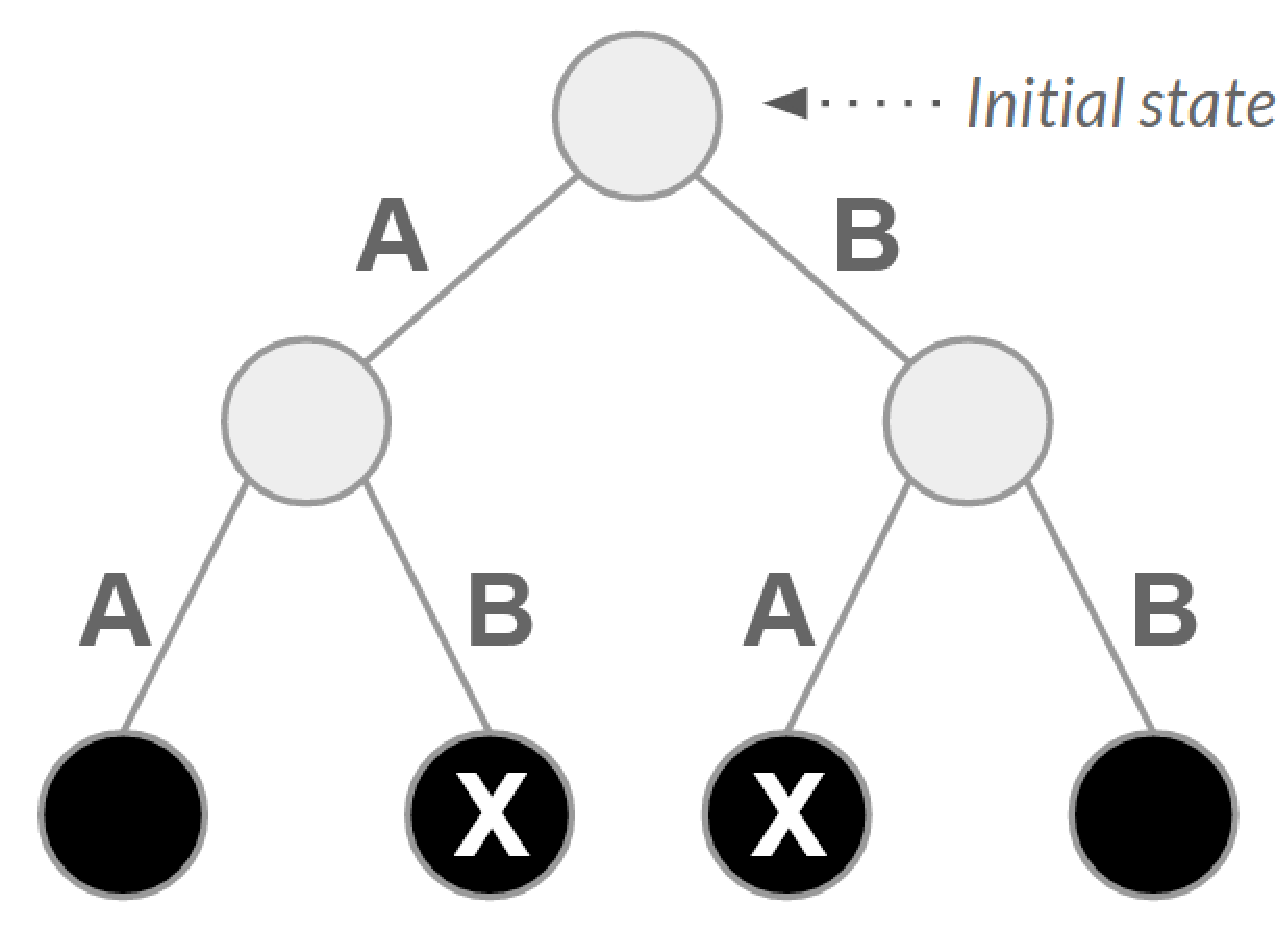}
\caption{{\bf Example of a state space.} Illustration of a full state space for a set of two \emph{reads}, here referred to by A and B.}
\label{fig:02}
\end{figure}

The reward system depends on the type of state reached after each action (absorbing or nonabsorbing). An absorbing state does not allow any other state~\cite{grinstead2012,sutton2018}. Each state requiring $n$ actions to be reached (with $n$ being the number of \emph{reads}) is an absorbing state. A small and constant reward (i.e., 0.1) is assigned for actions reaching nonabsorbing states, including actions leading to absorbing states achieved by repeated \emph{reads}. Finally, actions leading to other absorbing states (the final states) produce a reward corresponding to the sum of overlaps between all pairs of consecutive \emph{reads} used to reach these states. Fig \ref{fig:02} presents a state space example for a set of 2 \emph{reads} (A and B) with a single initial state, two actions associated with nonabsorbing states and four absorbing states (highlighted in black), achieved after taking two any actions.

Two absorbing states are highlighted ($X$). These are the final states, as they are reached directly without repeated actions. The Smith-Waterman algorithm (SW) was applied to obtain the overlaps between pairs of reads and added to obtain the rewards of actions leading to the final states~\cite{smith1981}. The sum of overlaps when reaching a final state $s$ (Performance Measure - PM), is described in Eq \ref{eq:03}, where $read_s$ correspond to the sequence of reads associated with the actions for achieving $s$. In an optimal solution repeated reads overlap completely and pairs reach the maximum PM.

\begin{equation}
\label{eq:03}
PM(s) = \sum_{i=1}^{n-1}sw(read_s[i], read_s[i+1])
\end{equation}

With these definitions the seminal approach produced positive results against two sets of 4 and 10 simulated \emph{reads} less than 10 $bp$ and 8 $bp$ respectively. A scalability analysis was applied to evaluate the performance of this approach against 18 datasets produced following the same simulation methods~\cite{xavier2020}. The initial set is one of the sets featured in the seminal approach, containing 10 \emph{reads} with 8 $bp$ extracted from a 25 $bp$ microgenome. Seventeen new datasets were generated from this microgenome and also from a novel 50 $bp$ microgenome (8 from the minor and 9 from the major microgenome) each containing 10, 20 or 30 \emph{reads}, with 8 $bp$, 10 $bp$ or 15 $bp$.

All previous definitions were replicated, but $\alpha$ and $\beta$ were set to $0.8$ and $0.9$, respectively, and the space of actions was reduced so that actions associated with previously taken \emph{reads} were removed from the available actions~\cite{xavier2020}. In the state space depicted in Fig \ref{fig:02} the leftmost and rightmost leaves (i.e., absorbing states) would be removed after this change. Although the number of states is reduced, the size of the state space grows exponentially, as in Eq \ref{eq:08}.

\begin{equation}
\label{eq:08}
\textrm{number of states}=\sum_{i=0}^{n}\frac{n!}{(n-i)!}
\end{equation}

This confirmed positive results from the seminal approach with the first dataset; however, the performance decreased with increasing size, reaching the target microgenome in only 2 out of the 17 major datasets. This may be related to the high cost required by the agent to explore a vast state space and to the failures in the reward system~\cite{xavier2020}. 

To investigate the application of reinforcement learning to genome assembly and address the current challenge of applying RL to real-world problems~\cite{dulac2019}, we analyzed the limits of RL to the Genome Assembly problem, a key problem for scientific development. We corrected previously described issues, explored the performance of an improved reward system and added complementary strategies to be incorporated into the seminal approach to obtain improved and automated genome assemblies via machine learning. 

 

\section*{Materials and methods}

\begin{figure}[!h]
\centering
\includegraphics[width=12cm,keepaspectratio]{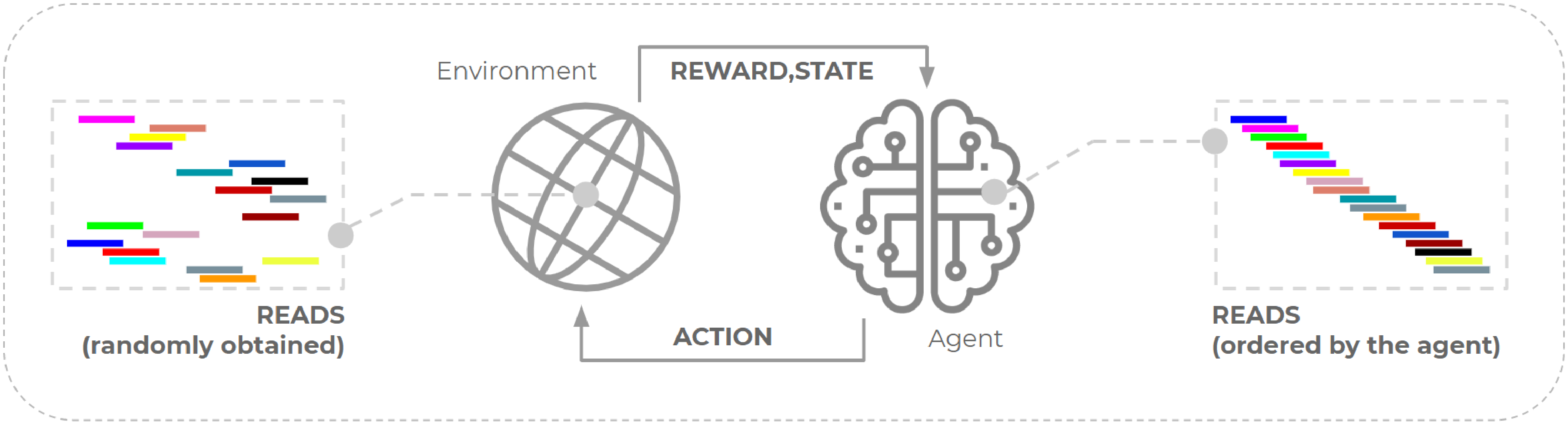}
\caption{{\bf Illustration of the application of reinforcement learning to the genome assembly problem.} The set of \emph {reads} is represented computationally by a reinforcement learning environment. Through successive interactions with the environment, caused by taking actions, the agent ideally learns the correct order of \emph {reads} --- reaching the target genome.}
\label{fig:07}
\end{figure}

In this study, 7 experiments were evaluated against the seminal approach. The main goal was to reach an RL trained agent, to correctly identify the order of \emph{reads} from a sequenced genome. Fig \ref{fig:07} illustrates this proposal, where the environment represents the set of \emph{reads} to assemble; the agent interacts with the environment by taking actions intended to order the \emph{reads}; for each action taken, the environment is updated and provides to the agent a corresponding reward; the agent learns from the reward received, and takes a new action, until reaching (ideally) the correct order of \emph{reads}.

The approaches produced here consider scalability analysis~\cite{xavier2020}, with improvements made to the reward system --- especially in Approaches 1  - and to optimize the agent's exploration --- approaches 2 and 3.

\subsection*{Approaches 1: Tackling sparse rewards}

Approaches 1.1, 1.2, 1.3, 1.4 aimed to improve the reward system, given by Eq \ref{eq:01}. Optimally, the agent achieves the correct order of \emph{reads} upon learning the set of actions, specifically a permutation of \emph{reads}, that maximizes the accumulated reward. Thus, the optimal actions -- those leading to the anticipated permutation of \emph{reads} -- must always yield the highest cumulative reward. Nevertheless, this proposition may not hold consistently for the reward system proposed in the seminal approach, as it allows some nonoptimal actions to result maximum accumulated rewards.

\begin{equation}
\label{eq:01}
r(s,a, s')=\left\{
\begin{array}{ll}
PM(s') & \mbox{if s' is a final state},\\
0.1 &\mbox{otherwise}
\end{array}
\right.
\end{equation}

This inconsistency (details in Section 6 of \nameref{S1_Appendix} in supporting information) stems from the sequences alignment with the \emph{Smith-Waterman} algorithm (SW), which calculates a score to represent major alignment size (even if partial), but has no constraint on the order between sequences. Thus the overlap score from the SW might induce the agent to find \emph{read} permutations with high overlap values in pairs of \emph{reads} without any \emph{suffix-prefix} alignment. So that, using PM score as a reward for training may be ineffective for some datasets.



Thus, aiming to improve the agent's performance, we adjusted the reward system through four approaches to explore two aspects: (a) the use of an overlap score that considers the relative order of \emph{reads} and/or (b) the use of dense rewards. These new reward systems are presented in approaches 1.1, 1.2, 1.3 and 1.4.

As in the seminal approach, approach 1.1 defines that actions leading to the final states produce a bonus reward (of 1.0), added to another numerical overlap score between all subsequent \emph{reads} used since the initial state. Thus, a reward corresponding to the sum of the normalized overlap score (ranging from 0 to 1) of each pair of \emph{reads} was produced considering their relative order. Every action leading to nonfinal states produces constant and low rewards (0.1). Eq \ref{eq:05} formalizes the reward system for Approach 1.1, with $PM_{norm}(s')$ representing the normalized overlap between the \emph{reads} used to reach $s'$ (details in Section 2 of \nameref{S1_Appendix} in supporting information).

\begin{equation}
\label{eq:05}
r(s,a, s')=\left\{
\begin{array}{ll}
PM_{norm}(s') + 1.0 & \mbox{if s' is a final state},\\
0.1 &\mbox{otherwise}
\end{array}
\right.
\end{equation}

Despite the overlap score considering the order of \emph{reads} in approach 1.1, it is susceptible to the sparse rewards problem --- as in the seminal approach. Although it often produces small, constant and positive rewards, and not a zero-value reward as applied in sparse reward systems, only few and sparse state-action pairs would produce higher rewards. In both systems (Eq \ref{eq:01} and \ref{eq:05}) no reward is provided during the learning process (since any \emph{read} incorporated would produce a reward of $0.1$). 

Thus, the agent's learning process depends exclusively on the sparse actions taken during the exploration of this state space, tending to take a long time due to the sparse rewards problem~\cite{trott2019}. In approaches 1.2, 1.3 and 1.4, we focused on improving it with higher rewards distributed for each action taken in each episode (previously obtained only at the end of the episode). These approaches focused on reducing or eliminating inconsistencies that allowed permutations of unaligned \emph{reads} to produce maximum accumulated rewards. Eq \ref{eq:07}, \ref{eq:06} and \ref{eq:02} represent the reward systems for approaches 1.2, 1.3 and 1.4, respectively, so that $ol_{norm}(s,s')$ represents the normalized overlap between two subsequent \emph{reads}.

\begin{equation}
\label{eq:07}
r(s,a, s')=PM_{norm}(s')
\end{equation}

\begin{equation}
\label{eq:06}
r(s,a, s')=\left\{
\begin{array}{ll}
PM_{norm}(s') + 1.0 & \mbox{if s' is a final state},\\
ol_{norm}(s, s') &\mbox{otherwise}
\end{array}
\right.
\end{equation}

\begin{equation}
\label{eq:02}
r(s,a, s')=\left\{
\begin{array}{ll}
ol_{norm}(s, s') + 1.0 & \mbox{if s' is a final state},\\
ol_{norm}(s, s') &\mbox{otherwise}
\end{array}
\right.
\end{equation}

\subsection*{Approach 2: Pruning-based elimination action}

\begin{figure}[!h]
\centering
\includegraphics[width=8cm,keepaspectratio]{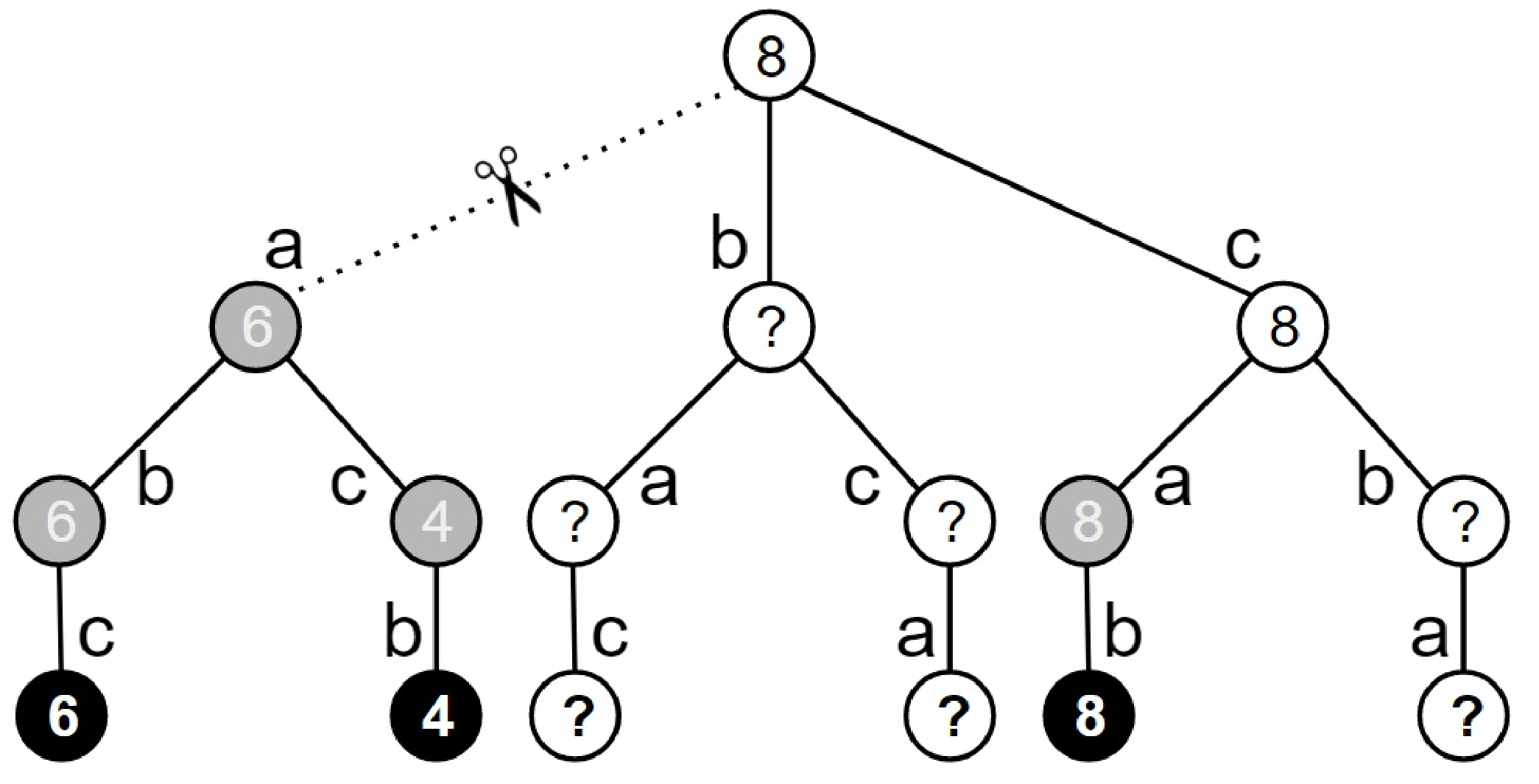}
\caption{{\bf Illustration of the pruning procedure.} State space corresponding to the assembly of 3 \emph{reads}, referred by \emph{a}, \emph{b} and \emph{c}. The generic pruning procedure is defined in detail by Algorithm \ref{alg:prunning}}
\label{fig:04}
\end{figure}

To reduce the state space from the seminal approach, a heuristic procedure was applied to eliminate fully explored actions where the maximum cumulative reward achieved was smaller than the cumulative reward from taking any other action available. In Fig \ref{fig:04}, looking at the changed state space as a tree --- removing actions associated with used \emph{reads}, we see 16 states, 6 are absorbing states and the final states (tree base). Note that 3 out of the 6 final states are highlighted in black, while the remaining states are highlighted in gray and white. The black states correspond to the explored final states (i.e., visited by the agent). Gray states, such as the one reached by taking action $a$ in the initial state, represent states where all children were fully visited during the learning process. White states (final or not) are those not yet explored and/or that have unexplored children --- e.g. the initial state, where one child is not explored and the other one is partially explored.

When an unexplored final state is reached, such as the rightmost final state in Fig \ref{fig:04}, the accumulated rewards are maintained and propagated for its predecessors, maintaining only the highest value propagated for the children. Each reward is represented by integer numbers within the states in the figure. In each nonfinal state the highest accumulated reward achieved during the training process is stored. Thus, it is possible to prune irrelevant actions, that do not produce the maximum accumulated reward (e.g., action \emph{a} of the initial state in Fig \ref{fig:04}). Note that all possible achievable states after taking this action were explored and the maximum cumulative reward was 6, while the initial state of action \emph{c} alone produces a reward equal to 8. When the agent first goes through the sequence of states corresponding to actions \emph{c}, \emph{a} and \emph{b}, the pruning mechanism propagates the maximum reward value up to the initial state and, at that moment, it cuts the action \emph{a} from the initial state. The pseudocode presented in Algorithm \ref{alg:prunning} presents the procedure for updating the pruning process when the last explored final state ($state$) is reached obtaining the corresponding accumulated reward achieved ($newReward$).

\begin{algorithm}

	\caption{Pruning’s algorithm}\label{alg:prunning}
	\begin{algorithmic}[1]
		\Procedure{Prune}{$state:treeNode,newReward:float$}

            \If{$state \not= null$ \textbf{and} $(state.unseen$ \textbf{or} $newReward > state.maxReward)$}
            \State $state.unseen \gets false$
            \State $state.maxReward \gets newReward$
            \If{$state.final$}\Comment{prune children where $maxReward < newReward$ }
            \State
            \textsc{PruneUselessChildren}($state$)
            
            \EndIf
            \State \textsc{Prune}($state.parent, newReward$)
            \EndIf

		\EndProcedure
	\end{algorithmic}
\end{algorithm}

\subsection*{Approaches 3: Evolutionary-based exploration}

In these approaches,  we explore the potential for mutual collaboration between reinforcement learning and evolutionary computing --– by applying the elitist selection of the genetic algorithm ~\cite{Baluja95, Konar2005} ---  to optimize the exploration of the state space. The individual contributions of the genetic algorithm used in this hybrid proposal were divided into two approaches, 3.1 and 3.2.

\subsubsection*{Approach 3.1: Evolutionary-aided reinforcement learning assembly}

\begin{figure}[!h]
\centering
\includegraphics[width=12cm,keepaspectratio]{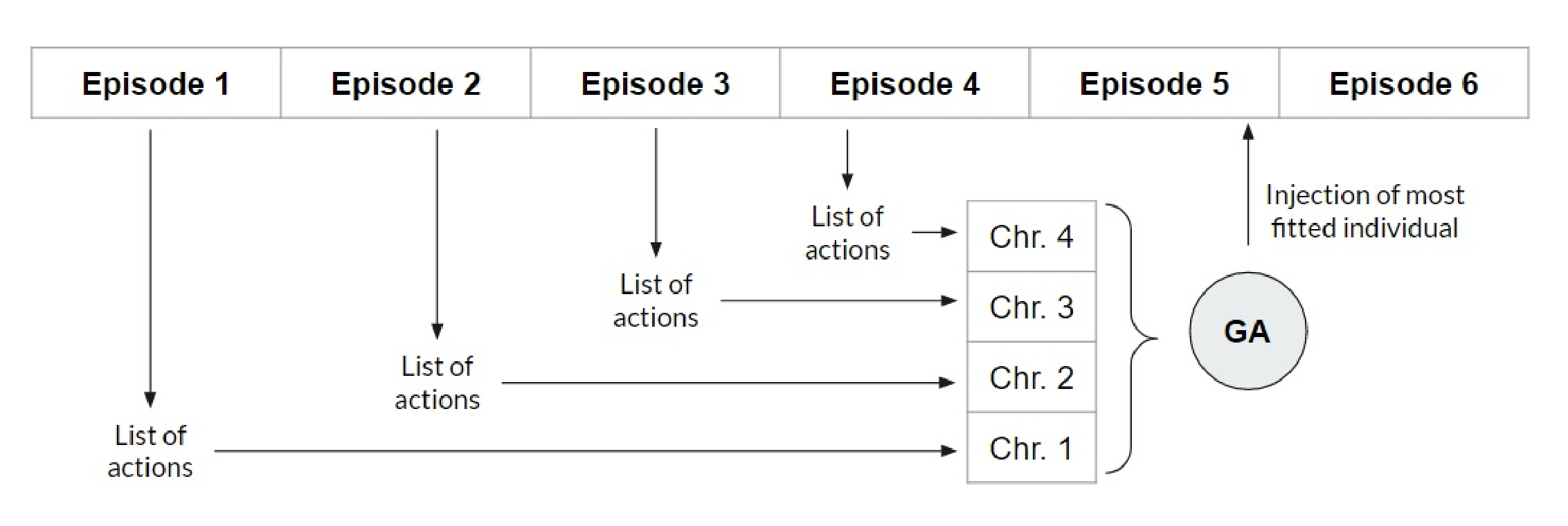}
\caption{{\bf Illustration of the proposed interaction between reinforcement learning (RL) and the genetic algorithm}. At each RL episode, the actions taken by the agent are converted into the chromosome (each action as a gene) of an individual of the initial population of the genetic algorithm, whose size $n$ is predefined. After $n$ episodes ($n$ individuals in the initial population), this population evolves for an predefined number of generations through the genetic algorithm. Then, the most adapted individual of the last generation is obtained. In the end, that individual's chromosomal genes are used as actions in the next RL episode.}
\label{fig:05}
\end{figure}

Applying the $\epsilon$-greedy to expand the exploration of agents trained by the Q-learning algorithm allows a broader initial exploration, achieving optimal policy once the state space has been sufficiently explored~\cite{sutton2018}. However, the existing trade-off between exploitation and exploration remains a major problem for RL in high-dimensional environments~\cite{gimelfarb2020, peterson2019}. Here, for the first time, we introduce the interaction between RL and evolutionary computing into the exploration process based on the operation of the Q-learning algorithm. In each episode, the sequence of actions is stored, and at the end of the episode the sequence is transformed into a chromosome of an initial population, that evolves (see Fig \ref{fig:05}).

New chromosomes are inserted until the number of chromosomes reaches the predefined population size. At this point, agent training is interrupted and $m$ genetic generations are carried out –-- with $m$ being predefined (see Section 4 of \nameref{S1_Appendix} in supporting information) and applying the normalized sum of overlaps between \emph{reads} as the adaptive function --- the same as that applied in Eq \ref{eq:02} and detailed in Section 2 of \nameref{S1_Appendix} in supporting information.

\begin{figure}[!h]
\centering
\includegraphics[width=12cm,keepaspectratio]{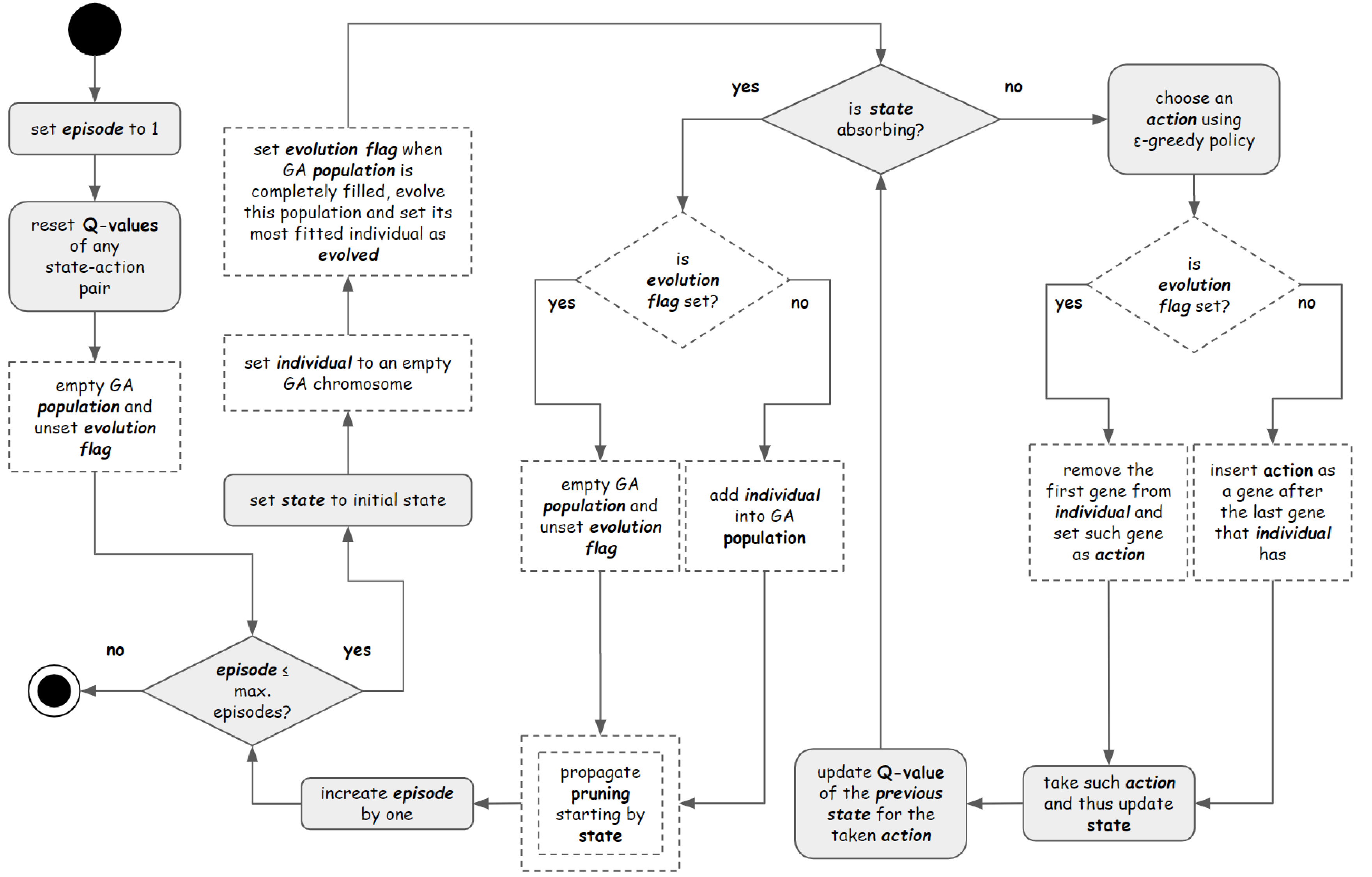}
\caption{{\bf Flowchart representing Approaches 1.1, 1.2, 1.3, 1.4, 2 and 3.1.} Approaches 1.1, 1.2, 1.3, and 1.4 are defined by the elements in gray, Approach 2 by the dashed element with double edges and Approach 3.1 by the dashed elements with single border.}
\label{fig:06}
\end{figure}

After $m$ generations (objective function) the most fit individual is used for conducting the next episode in the agent's RL training, hitherto interrupted. As each gene of the individual's chromosome corresponds to one possible action, the complete gene sequence will contain distinct successive actions to be taken by the agent in the current episode, producing a mutual collaboration between RL and the genetic algorithm --- the initial populations of the genetic algorithm are produced by RL and, as a counterpart, the results from the evolution of the genetic algorithm are introduced in an RL episode (Fig \ref{fig:06}).

\subsubsection*{Approach 3.2: Evolutionary-based assembly}

To estimate the genetic algorithm contribution in Approach 3.1, its assembling performance was evaluated separately, following the same configurations set for the previous approach, but adopting as a starting population a set of individuals whose chromosomes were built from random permutations without repetition of \emph{reads}.

\subsection*{Datasets and analysis}

To assess the performance of all the approaches (including the seminal approach), in addition to the 18 datasets from Xavier et al.~\cite{xavier2020}, 5 novel datasets derived from microgenomes extracted in previous studies~\cite{bocicor2011, xavier2020} were created. These are not arbitrary genome fragments, as were the case for previously used microgenomes (which had 25 $bp$ and 50 $bp$), but represent larger fragments of previously annotated genes from the corresponding organism (i.e., \emph{E. coli}). Given that the datasets are simulated data, no cycles in the genome were considered, which is a limitation of the approach. The experiments were carried out with 23 datasets, detailed in Table \ref{tab:03} --- the last 5 lines are gene derived.

An environment for each dataset was created in the OpenAI Gym toolkit~\cite{brockman2016} to share such RL challenges. These environments are available online (see Section 1 of \nameref{S1_Appendix} in supporting information), where the reward system proposed in Approach 1.4 is used. The identification names of each environment are presented in the last column of Table \ref{tab:03}. The seminal reward system is also implemented and available –-- version 1, replacing \emph{v2} with \emph{v1} in the environment name field.

\begin{table}[!ht]
\centering
\caption{\label{tab:03} {\bf Public datasets used in the experiments.}}
\begin{tabular}{|c|c|c|c|}
\hline
\textbf{$\mu$gen.} & \textbf{\#} & \textbf{read} & \textbf{Gym} \\
\textbf{size} & \textbf{reads} & \textbf{size} & \textbf{environment name} \\
\thickhline
25 & 10 & 8 & GymnomeAssembly\_25\_10\_8-v2 \\ \hline
25 & 10 & 10 & GymnomeAssembly\_25\_10\_10-v2 \\ \hline
25 & 10 & 15 & GymnomeAssembly\_25\_10\_15-v2 \\ \hline
50 & 10 & 8 & GymnomeAssembly\_50\_10\_8-v2 \\ \hline
50 & 10 & 10 & GymnomeAssembly\_50\_10\_10-v2 \\ \hline
50 & 10 & 15 & GymnomeAssembly\_50\_10\_15-v2 \\ \hline
25 & 20 & 8 & GymnomeAssembly\_25\_20\_8-v2 \\ \hline
25 & 20 & 10 & GymnomeAssembly\_25\_20\_10-v2 \\ \hline
25 & 20 & 15 & GymnomeAssembly\_25\_20\_15-v2 \\ \hline
50 & 20 & 8 & GymnomeAssembly\_50\_20\_8-v2 \\ \hline
50 & 20 & 10 & GymnomeAssembly\_50\_20\_10-v2 \\ \hline
50 & 20 & 15 & GymnomeAssembly\_50\_20\_15-v2 \\ \hline
25 & 30 & 8 & GymnomeAssembly\_25\_30\_8-v2 \\ \hline
25 & 30 & 10 & GymnomeAssembly\_25\_30\_10-v2 \\ \hline
25 & 30 & 15 & GymnomeAssembly\_25\_30\_15-v2 \\ \hline
50 & 30 & 8 & GymnomeAssembly\_50\_30\_8-v2 \\ \hline
50 & 30 & 10 & GymnomeAssembly\_50\_30\_10-v2 \\ \hline
50 & 30 & 15 & GymnomeAssembly\_50\_30\_15-v2 \\ \hline
381 & 20 & 75 & GymnomeAssembly\_381\_20\_75-v2 \\  \hline
567 & 30 & 75 & GymnomeAssembly\_567\_30\_75-v2 \\  \hline
726 & 40 & 75 & GymnomeAssembly\_728\_40\_75-v2 \\  \hline
930 & 50 & 75 & GymnomeAssembly\_930\_50\_75-v2 \\  \hline
4224 & 230 & 75 & GymnomeAssembly\_4224\_230\_75-v2 \\ \hline
\end{tabular}
\begin{flushleft} The first column shows the size (in $bp$) of the microgenome used to generate the \emph{reads} of each set; the second column shows the number of \emph{reads} generated; the third column shows the size of the generated \emph{reads}; and the fourth column shows the name of the environment built for each set in the OpenAI Gym toolkit.
\end{flushleft}
\end{table}

Two experiments were carried out to evaluate the approaches. In each experiment, 20 successive runs of each evaluated approach were performed for all 23 existing datasets (460 runs per approach). Given that each approach has different levels of complexity, the real execution time for each approach was considered for comparison. To reduce the interference of external factors in execution time, all experiments were individually and sequentially performed at the same station (with Ubuntu 16.04 in an AWS EC2 instance of the \emph{r5a.large} type, dual core, 16 GB of RAM and 30 GB of storage).

In the first experiment (hereinafter referred to as experiment A) the objective was to verify the impact of progressively including new strategies. For this purpose, the performance of the seminal approach was evaluated (according to \cite{bocicor2011}) against approaches 1.1, 1.2, 1.3, 1.4 (improved reward system), 2 (pruning dynamic) and 3.1 (genetic algorithm - GA). In the second experiment (or experiment B) the objective was to compare the performance of the new RL-based approaches against the performance of the GA alone. Therefore, in addition to Approaches 1.1, 1.2, 1.3, 1.4, 2 and 3.1, the approach 3.2 (which explores GA alone) was performed in an equivalent time.

For the performance measure in each experiment, two percentage measures were calculated, called the distance-based measure (DM) and reward-based measure (RM). Evaluations of \emph{de novo} assembly are commonly performed using proper metrics, such as the N50~\cite{Bradnam2013}. These metrics were created because, as previously indicated, \emph{de novo} assemblies are not supported by a reference genome. In some scenarios, it is not possible to accurately assess the results obtained from the assemblers –-- because the optimal output is unknown. Here, although a \emph{de novo} assembler is evaluated, its assessment environment is restricted, and the target genomes are known; this scenario allows the use of specific (and exact) evaluations, such as DM and RM metrics. 

DM considers a successful run when the consensus sequence from the orders of \emph{reads} produced is identical to the expected sequence. RM considers any run as a success when the proposed order of \emph{reads} represents the sum of PM$_{norm}$ higher than or equal to the sum of PM$_{norm}$ from the optimal \emph{read} sequence (for details, see Section 3 of \nameref{S1_Appendix} in supporting information).

\section*{Results}

In experiment A the seminal approach consumed the longest running time (23 hours and 34 minutes)and had the lowest average performance; an optimal response was obtained in 16.96\% of the runs (i.e., 78 out of the 460 executions) in terms of distance from the expected genome (DM) and 21.30\% (98 out of the 460) in terms of maximum reward (RM) (Table \ref{tab:01}). 

\begin{table}[!ht]
\centering
\caption{\label{tab:01}{\bf Results achieved in experiment A.}}
\begin{tabular}{|c|c|c|c|}
\hline
\textbf{Experiment A} & \textbf{Average} & \textbf{Average} &\textbf{Total} \\
\textbf{(Approach)} & \textbf{DM} & \textbf{RM} &\textbf{runtime} \\ \thickhline
Seminal & 16.96\% & 21.30\% & 23h34m \\ \hline
1.1 & 9.57\% & 13.70\% & 19h38m \\ \hline
1.2 & 18.48\% & 21.30\% & 19h38m \\ \hline
1.3 & 20.00\% & 24.35\% & 19h38m \\ \hline
1.4 & 20.43\% & 24.78\% & 19h38m \\ \hline
2 & 20.65\% & 25.00\% & 18h41m \\ \hline
3.1 & \textbf{73.91\%} & \textbf{80.87\%} & 17h03m \\ \hline
\end{tabular}
\begin{flushleft}The performances of each approach are expressed using distance-based (DM) and reward-based (RM) metrics (see \emph{Methods} for details).
\end{flushleft}
\end{table}

Following the updated reward system, DM and MR performances in Approaches 1.2, 1.,3, and 1.4 surpassed those of the previous approach, and consumed approximately 4 hours less (19 hours and 38 minutes) of running time. Approach 3.1 presented the shortest running time, with a DM average of approximately 74\% and an RM average above 80\%, with the highest performance.  

In experiment B, according to Table \ref{tab:02}, approach 3.2 presented the shortest running time, with a DM average of 87\% and an RM average of 95\%. Given the superior performance of Approach 3.2, Experiment B applied the time taken by the Genetic Algorithm as a reference to find an optimal solution in terms of the RM for 22 out of the the 23 datasets used (i.e., $95.65\%$), which corresponded to 1 hour and 34 minutes of running time. Given the dominance of Approach 3.2, we also verified the performance of this approach on only the dataset with no optimal response (reads with 4Kbp). 

\begin{table}[!ht]
\centering
\caption{\label{tab:02}{\bf Experimental performances considering similar running times (RT)}}.
\begin{tabular}{|c|c|c|c|}
\hline
\textbf{Experiment B} & \textbf{Average} & \textbf{Average} &\textbf{Total} \\
\textbf{(Approach)} & \textbf{DM} & \textbf{RM} &\textbf{runtime} \\ \thickhline
1.4 & 13.91\% & 17.61\% & 01h36m \\ \hline
2 & 12.39\% & 16.30\% & 01h36m \\ \hline
3.1 & 14.78\% & 14.78\% & 01h42m \\ \hline
3.2 & \textbf{87.83\%} & \textbf{95.65\%} & 01h34m \\\hline

\end{tabular}
\begin{flushleft}
Performances were expressed using Distance-based Measure (DM) and Reward-based Measure (RM) (see \emph{Methods}).
\end{flushleft}
\end{table}

\begin{figure}[!ht]
\centering
\includegraphics[width=\linewidth]{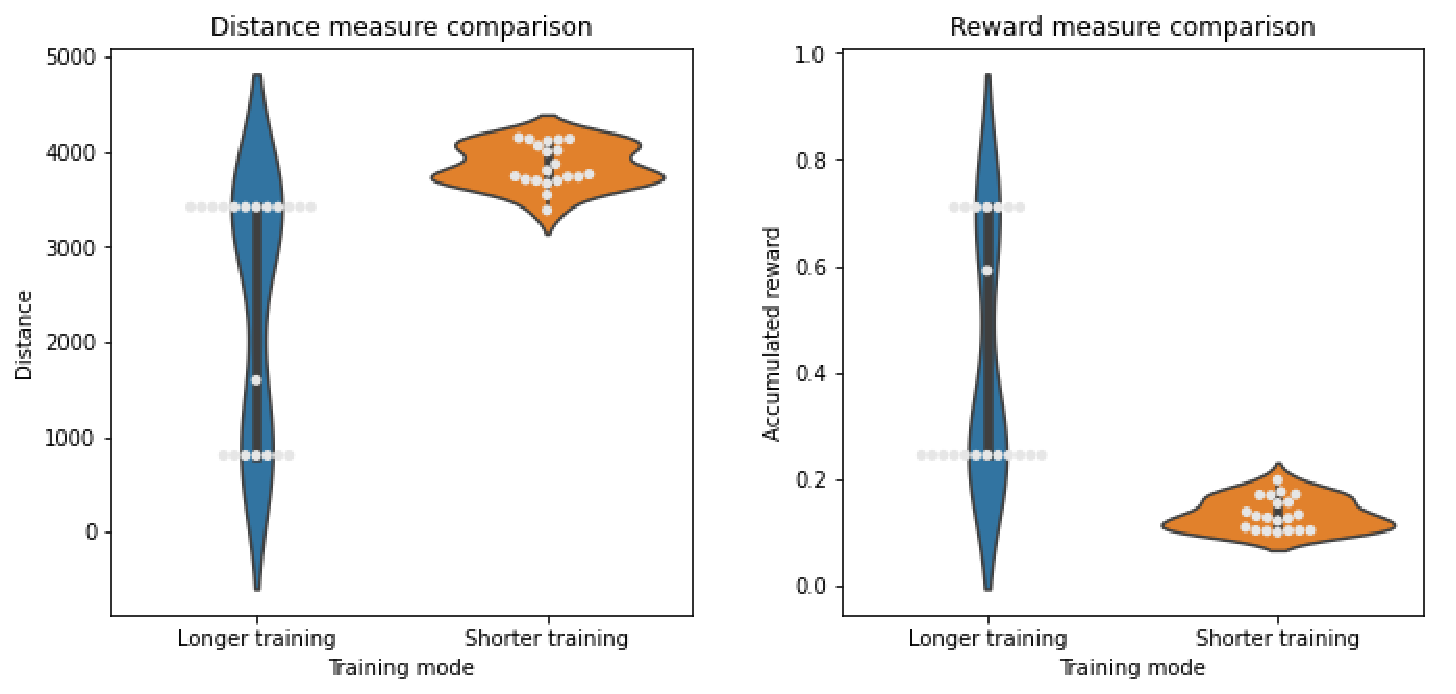}
\caption{{\bf GA performance evaluation on a long run.} Violin plots demonstrating the performances obtained by the GA in experiments with short (1 h 34 m) and long (37 h 58 m) execution times in terms of the sums of DM and RM. The gray dots represent the distances/rewards obtained for all runs; the black line in the middle indicates the interquartile range; and the violin curves show the distribution density, where the wider the section, the greater the probability of the observations taking the corresponding value.}
\label{fig:01}
\end{figure}

In this experiment the running time for Approach 3.2 was considerably increased running time, lasting approximately 38 hours  (against less than 2 minutes for the same dataset for approach 3.2 in experiment B). No optimal solution was obtained for this dataset, however, it is possible to observe a consistent gain in performance, in terms of both DM (where longer runs had shorter distances than most distances reached by shorter runs) and RM (which had higher accumulated rewards in all longer runs) (Fig \ref{fig:01}).

All the data generated or analyzed during this study, including reproduction codes, are also publicly available (see Section 5 of \nameref{S1_Appendix} in supporting information).

\section*{Discussion}

Genome assembly is among the most complex problems confronted by computer scientists within the context of genomics projects. When applying machine learning to genome assembly this complexity allocates the problem of finding optimal permutations of sequenced \emph{reads} and reaching the target genome into an NP-hard problem, which comprises the most difficult problems in computer science~\cite{roughgarden2020}. This high complexity is particularly expressed in the vast state space required for representing the assembly problem in RL models. To achieve the optimal solution in sets of 30 \emph{reads} the RL agent should explore a state space of approximately $2e^{44}$ states~\cite{bocicor2011} (more than the stars in the universe). In real-world scenarios genomes would be much larger. Applying RL combined with heuristics is a strategy for addressing complex problems, aiming at mapping actions into states that tend to maximize their reward, thus decreasing the computational complexity of the problem.

In this study, we aimed to expand agent learning based on two difficulties observed in the seminal approach for applying RL to the genome assembly problem: (1) the reward system and (2) the agent's exploration strategies. We found that both improving the agent's learning performance and updating the reward systems favored the agent to improve learning. However, this was not yet an optimal solution, as nonoptimal solutions were still present (less frequent) in the reward systems. This is also supported by the fact that RM percentages were higher than DM percentages in some experiments. 

The dynamic pruning mechanism showed slight improvement, but the additional processing cost and the benefit from its implementation did not indicate a reasonable net gain from its use as bypass for the problem emerging from the high dimensionality of the state space. Part of the gains were due to the improved agent's performance where the sum of rewards for the optimal permutation of \emph{reads} was not maximized in the previous reward system. Despite the gains from the updated reward system, the inconsistencies were not completely resolved. In some of the datasets the agent reached and even surpassed the maximum expected accumulated rewards without obtaining the target genome. A minor improvement is observed in approach 2, requiring approximately one hour less of processing.

The hybrid approach combining RL strategy with GA (Approaches 3), presented better performances. This combination was proven to be advantageous, probably given the curse of dimensionality  encountered by the Q-learning algorithm, as a strong GA support was observed for the agent while conducting the RL exploration. 

Despite these improvements, the approaches are not suitable for real-world scenarios. This is evident in the experiments performed with the largest dataset. Even the smallest genomes from living organisms is greater than the largest dataset of this study and none of the proposed approaches produced an optimal response for this dataset, even when the best approach (GA) was applied for a longer time. 

The superiority of the GA alone allows us to conclude on the infeasibility of applying the Q-learning algorithm to solve the genome assembly problem in search of an optimal \emph{read} permutation, as proposed in the seminal approach. 

Given the absence of approaches in the literature for tackling this problem through RL and considering the optimistic results obtained by RL in other areas (especially when RL is combined to deep learning), further investigations on the applicability of RL, including the use of different modeling approaches and algorithms, are needed.

All the experiments and the RL environments used in this study, are publicly available and open for reuse (for details see Section 5 of \nameref{S1_Appendix} in supporting information), to support future studies.

One of the major challenges in applying RL to real-world problems is the low sample efficiency of the algorithms~\cite{yang2018}. Considering the time required by the agent trained by the Q-learning algorithm to reach an optimal solution, it is possible to perceive a high need for numerous interactions with the data. Considering that genome inputs are larger than those experimentally applied here, obtaining a sample efficient algorithm for the problem is in the core for developing a real-world solution. Additionally, the agent sample efficiency must be optimized to explore the state space, which might be achieved by the application of techniques to remove duplicate \emph{reads} --- due to repeats --- and the use of an intrinsic motivation to bypass the exploration problem, given the high dimensionality of the proposed state space~\cite{barto2012, yang2018}.

Although RL faces obstacles for commercial application, as mentioned before, it seems reasonable to consider the recent achievements of deep RL when applied to games, which present an equivalent computational problem to that of genome assembly~\cite{Vinyals2019,Silver2017,fjelland2020}. The transformation of real-world problems into games is also a possibility for applying developed technologies ~\cite{Reis2020}. One of the main benefits of representing the problem as a game is the reduction of the space of actions, which increases with the number of \emph{reads} in the approaches proposed here.

The use of \emph{graph embedding} may act as another option for modeling approaches allowing the use of deep RL without requiring the conversion of the problem into an image –- the genome assembly problem may be represented through a graph, in the shape of the traveling salesman problem (TSP)~\cite{Cook2012, Zhenyu2011}. 

Finally, one last aspect to be considered for the adoption of RL into the genome assembly problem is the generalization of the agent´s learning -- a major challenge for the use of RL in real-world problems~\cite{Ponsen2010}. As designed for the RL environment for the genome assembly problem, the learning acquired by the agent when assembling a set of \emph{reads} will hardly be applied for the assembly of a new set.

\section*{Supporting information}

\paragraph*{S1 Appendix.}
\label{S1_Appendix}
{\bf Supplementary Material.} This is the supplementary material associated to this manuscript, containing further explanations.

\paragraph*{S1 Fig.}
\label{S1_Fig}
{\bf Example of a state space.} Image file containing the picture corresponding to Fig \ref{fig:02}.

\paragraph*{S2 Fig.}
\label{S2_Fig}
{\bf Illustration of the application of reinforcement learning to the genome assembly problem.} Image file containing the picture corresponding to Fig \ref{fig:07}.

\paragraph*{S3 Fig.}
\label{S3_Fig}
{\bf Illustration of the pruning procedure.} Image file containing the picture corresponding to Fig \ref{fig:04}.

\paragraph*{S4 Fig.}
\label{S4_Fig}
{\bf Illustration of the proposed interaction between reinforcement learning (RL) and the genetic algorithm.} Image file containing the picture corresponding to Fig \ref{fig:05}.

\paragraph*{S5 Fig.}
\label{S5_Fig}
{\bf Flowchart representing Approaches 1.1, 1.2, 1.3, 1.4, 2 and 3.1.} Image file containing the picture corresponding to Fig \ref{fig:06}.

\paragraph*{S6 Fig.}
\label{S6_Fig}
{\bf GA performance evaluation on a long run.} Image file containing the picture corresponding to Fig \ref{fig:01}.






\nolinenumbers

%
%
%




\bibliography{main}

\begin{thebibliography}{10}

\bibitem{Portin2017}
Portin P, Wilkins A.
\newblock The Evolving Definition of the Term {\textquotedblleft}Gene{\textquotedblright}.
\newblock Genetics. 2017;205(4):1353--1364.
\newblock doi:{10.1534/genetics.116.196956}.

\bibitem{Heather2016}
Heather JM, Chain B.
\newblock The sequence of sequencers: The history of sequencing {DNA}.
\newblock Genomics. 2016;107(1):1--8.
\newblock doi:{10.1016/j.ygeno.2015.11.003}.

\bibitem{Ji2017}
Ji P, Zhang Y, Wang J, Zhao F.
\newblock {MetaSort} untangles metagenome assembly by reducing microbial community complexity.
\newblock Nature Communications. 2017;8(1).
\newblock doi:{10.1038/ncomms14306}.

\bibitem{Wong2020}
Wong HL, MacLeod FI, White RA, Visscher PT, Burns BP.
\newblock Microbial dark matter filling the niche in hypersaline microbial mats.
\newblock Microbiome. 2020;8(1).
\newblock doi:{10.1186/s40168-020-00910-0}.

\bibitem{Medvedev2007}
Medvedev P, Georgiou K, Myers G, Brudno M.
\newblock Computability of Models for Sequence Assembly.
\newblock In: Lecture Notes in Computer Science; 2007. p. 289--301.

\bibitem{Gurevich2013}
Gurevich A, Saveliev V, Vyahhi N, Tesler G.
\newblock {QUAST}: quality assessment tool for genome assemblies.
\newblock Bioinformatics. 2013;29(8):1072--1075.
\newblock doi:{10.1093/bioinformatics/btt086}.

\bibitem{PadovanideSouza2018}
de~Souza KP, Setubal JC, de~Leon F~de Carvalho ACP, Oliveira G, Chateau A, Alves R.
\newblock Machine learning meets genome assembly.
\newblock Brief in Bioinformatics. 2018;doi:{10.1093/bib/bby072}.

\bibitem{Yassine2023}
Yassine A, Riffi ME.
\newblock A Review on Machine-Learning and Nature-Inspired Algorithms for Genome Assembly.
\newblock International Journal of Advanced Computer Science and Applications (IJACSA). 2023;14(7):898.
\newblock doi:{10.14569/issn.2156-5570}.

\bibitem{LeCun2019}
LeCun Y.
\newblock Deep Learning Hardware: Past, Present, and Future.
\newblock In: 2019 {IEEE} International Solid- State Circuits Conference - ({ISSCC}); 2019.

\bibitem{Botvinick2019}
Botvinick M, Ritter S, Wang JX, Kurth-Nelson Z, Blundell C, Hassabis D.
\newblock Reinforcement Learning, Fast and Slow.
\newblock Trends in Cognitive Sciences. 2019;23(5):408--422.
\newblock doi:{10.1016/j.tics.2019.02.006}.

\bibitem{sutton2018}
Sutton RS, Barto AG.
\newblock Reinforcement Learning: An Introduction.
\newblock Cambridge, MA, USA: A Bradford Book; 2018.
\newblock Available from: \url{http://incompleteideas.net/book/RLbook2020.pdf}.

\bibitem{bocicor2011}
Bocicor MI, Czibula G, Czibula IG.
\newblock A Reinforcement Learning Approach for Solving the Fragment Assembly Problem.
\newblock In: 2011 13th International Symposium on Symbolic and Numeric Algorithms for Scientific Computing; 2011.

\bibitem{xavier2020}
Xavier R, de~Souza KP, Chateau A, Alves R.
\newblock Genome Assembly Using Reinforcement Learning.
\newblock In: Kowada L, de~Oliveira D, editors. Advances in Bioinformatics and Computational Biology". Cham: Springer International Publishing; 2020. p. 16--28.

\bibitem{Karami2023}
Karami M, Alizadehsani R, Jahanian K, Argha A, Dehzangi I, Alinejad-Rokny H. Revolutionizing Genomics with Reinforcement Learning Techniques; 2023.

\bibitem{cormen2009}
Cormen TH, Leiserson CE, Rivest RL, Stein C.
\newblock Introduction to Algorithms, Third Edition.
\newblock Cambridge, MA, USA: The MIT Press; 2009.

\bibitem{grinstead2012}
Grinstead CM, Snell JL.
\newblock Introduction to Probability; 2012.
\newblock Available from: \url{https://books.google.com.br/books?id=7ip55ODL72wC}.

\bibitem{smith1981}
Smith TF, Waterman MS.
\newblock Identification of common molecular subsequences.
\newblock Journal of Molecular Biology. 1981;147(1):195--197.
\newblock doi:{10.1016/0022-2836(81)90087-5}.

\bibitem{dulac2019}
Dulac{-}Arnold G, Mankowitz DJ, Hester T.
\newblock Challenges of Real-World Reinforcement Learning.
\newblock In: ICML 2019 Workshop on Reinforcement Learning for Real Life (RLRL); 2019.

\bibitem{trott2019}
Trott A, Zheng S, Xiong C, Socher R.
\newblock Keeping Your Distance: Solving Sparse Reward Tasks Using Self-Balancing Shaped Rewards.
\newblock In: Wallach HM, Larochelle H, Beygelzimer A, d'Alch{\'{e}}{-}Buc F, Fox EB, Garnett R, editors. Advances in Neural Information Processing Systems 32: Annual Conference on Neural Information Processing Systems 2019, NeurIPS 2019, 8-14 December 2019, Vancouver, BC, Canada; 2019. p. 10376--10386.

\bibitem{Baluja95}
Baluja S, Caruana R.
\newblock Removing The Genetics from The Standard Genetic Algorithm.
\newblock In: In Proceedings of ICML’95. California: Elsevier; 1995. p. 38--46.

\bibitem{Konar2005}
Konar A.
\newblock Evolutionary Computing Algorithms.
\newblock In: Computational Intelligence; 2005. p. 323--351.

\bibitem{gimelfarb2020}
Gimelfarb M, Sanner S, Lee CG.
\newblock Epsilon-BMC: A Bayesian Ensemble Approach to Epsilon-Greedy Exploration in Model-Free Reinforcement Learning.
\newblock In: Adams RP, Gogate V, editors. Proceedings of Machine Learning Research. vol. 115. Tel Aviv, Israel: PMLR; 2020. p. 476--485.

\bibitem{peterson2019}
Peterson EJ, Verstynen TD.
\newblock A way around the exploration-exploitation dilemma.
\newblock bioRxiv. 2019;doi:{10.1101/671362}.

\bibitem{brockman2016}
Brockman G, Cheung V, Pettersson L, Schneider J, Schulman J, Tang J, et~al.. OpenAI Gym; 2016.

\bibitem{Bradnam2013}
Bradnam KR, Fass JN, Alexandrov A, Baranay P, Bechner M, Birol I, et~al.
\newblock Assemblathon 2: evaluating de novo methods of genome assembly in three vertebrate species.
\newblock {GigaScience}. 2013;2(1).
\newblock doi:{10.1186/2047-217x-2-10}.

\bibitem{roughgarden2020}
Roughgarden T.
\newblock Algorithms Illuminated (Part 4): Algorithms for NP-Hard Problems.
\newblock Algorithms Illuminated; 2020.
\newblock Available from: \url{https://books.google.com.br/books?id=FlmuzQEACAAJ}.

\bibitem{yang2018}
Yu Y.
\newblock Towards Sample Efficient Reinforcement Learning.
\newblock In: Proceedings of the 27th International Joint Conference on Artificial Intelligence. IJCAI'18; 2018. p. 5739–5743.

\bibitem{barto2012}
Barto AG.
\newblock Intrinsic Motivation and Reinforcement Learning.
\newblock In: Intrinsically Motivated Learning in Natural and Artificial Systems; 2012. p. 17--47.

\bibitem{Vinyals2019}
Vinyals O, Babuschkin I, Czarnecki WM, Mathieu M, Dudzik A, Chung J, et~al.
\newblock Grandmaster level in {StarCraft} {II} using multi-agent reinforcement learning.
\newblock Nature. 2019;575(7782):350--354.
\newblock doi:{10.1038/s41586-019-1724-z}.

\bibitem{Silver2017}
Silver D, Schrittwieser J, Simonyan K, Antonoglou I, Huang A, Guez A, et~al.
\newblock Mastering the game of Go without human knowledge.
\newblock Nature. 2017;550(7676):354--359.
\newblock doi:{10.1038/nature24270}.

\bibitem{fjelland2020}
Fjelland R.
\newblock Why general artificial intelligence will not be realized.
\newblock Humanities and Social Sciences Communications. 2020;7(1).
\newblock doi:{10.1057/s41599-020-0494-4}.

\bibitem{Reis2020}
Reis S, Reis LP, Lau N.
\newblock Game Adaptation by Using Reinforcement Learning Over Meta Games.
\newblock Group Decision and Negotiation. 2020;doi:{10.1007/s10726-020-09652-8}.

\bibitem{Cook2012}
Cook WJ.
\newblock In: Pushing the Limits; 2012. p. 211--212.
\newblock Available from: \url{http://www.jstor.org/stable/j.ctt7t8kc.15}.

\bibitem{Zhenyu2011}
Li Z, Chen Y, Mu D, Yuan J, Shi Y, Zhang H, et~al.
\newblock {Comparison of the two major classes of assembly algorithms: overlap–layout–consensus and de-bruijn-graph}.
\newblock Briefings in Functional Genomics. 2011;11(1):25--37.
\newblock doi:{10.1093/bfgp/elr035}.

\bibitem{Ponsen2010}
Ponsen M, Taylor ME, Tuyls K.
\newblock Abstraction and Generalization in Reinforcement Learning: A Summary and Framework.
\newblock In: Adaptive and Learning Agents; 2010. p. 1--32.

\end{thebibliography}

\end{document}